\documentclass[12pt]{iopart}

\usepackage{iopams}
\usepackage{graphicx}
\begin{document}

\title[Asymptotic Exponents from Low-Reynolds-Number Flows]
{Asymptotic Exponents from Low-Reynolds-Number Flows}

\author{J\"org Schumacher$^1$, Katepalli R Sreenivasan$^2$ and Victor Yakhot$^3$}

\address{$^1$ Department of Mechanical Engineering, Technische Universit\"at Ilmenau,
         D-98684 Ilmenau, Germany}
\ead{joerg.schumacher@tu-ilmenau.de}

\address{$^2$ International Centre for Theoretical Physics, 34014 Trieste, Italy}
\ead{krs@ictp.it}

\address{$^3$ Department of Aerospace and Mechanical Engineering, Boston University,
         Boston, MA 02215, USA}
\ead{vy@bu.edu}

\begin{abstract}
The high-order statistics of fluctuations in velocity gradients in
the crossover range from the inertial to the Kolmogorov and
sub-Kolmogorov scales are studied by direct numerical simulations
(DNS) of homogeneous isotropic turbulence with vastly improved
resolution. The derivative moments for orders $0\le n\le 8$ are
represented well as powers of the Reynolds number, $Re$, in the
range $380\le Re\le 5725$, where $Re$ is based on the periodic box
length $L_x$. These {\em low-Reynolds-number flows} give no hint of
scaling in the inertial range even when extended self-similarity is
applied. Yet, the DNS scaling exponents of velocity gradients agree
well with those deduced, using a recent theory of anomalous scaling,
from the scaling exponents of the longitudinal structure functions
at {\em infinitely high Reynolds numbers}. This {\it suggests} that
the asymptotic state of turbulence is attained for the velocity
gradients at far lower Reynolds numbers than those required for the
inertial range to appear. We discuss these findings in the light of
multifractal formalism. Our numerical studies also resolve the
crossover of the velocity gradient statistics from the Gaussian to
non-Gaussian behaviour that occurs as the Reynolds number is
increased.
\end{abstract}

\submitto{\NJP}
%\tableofcontents
\maketitle

\section{Introduction}
\subsection{Motivation and Previous Work}
A deep understanding of the turbulent flow field ${\bf u}({\bf
x},t)$ remains a challenging problem. Extensions of the classical
theory of turbulence by Kolmogorov \cite{k41} consider a
multiplicity of algebraic scaling exponents for moments of velocity
increments $\delta_{r}u$ in the inertial range of length scales $r$,
which are spanned roughly between the Kolmogorov dissipation scale
$\eta_K$ and the outer scale of turbulence $L$. The longitudinal increment
moments (or structure functions) are then given as
\begin{equation}
S_{n}(r)\equiv \overline{(\delta_{r} u)^{n}}= \overline{\left(({\bf
u}({\bf x}+{\bf r})-{\bf u}({\bf x})){\bf\cdot}\frac{\bf
r}{r}\right)^n} =A_n\left(\frac{r}{L}\right)^{\zeta_n}\,,
\label{structuref}
\end{equation}
where the scaling exponents $\zeta_n$ depend nonlinearly on the
order $n$ but not on the Reynolds number $Re$, as long as the latter
is sufficiently large. The dimensional coefficients $A_n$ depend at
most on large-scale quantities. This nonlinear dependence of the
algebraic scaling exponents $\zeta_n$ on the moment order $n$ is a
manifestation of the inertial-range intermittency, which is
generally agreed to be an important feature of three-dimensional
turbulence. Inertial-range intermittency was experimentally first
quantified by Anselmet {\it et al.} \cite{Anselmet1984}. Starting
with the work of Kolmogorov \cite{k62} and Oboukhov
\cite{Obukhov1962}, numerous phenomenological models have been
developed to study and describe intermittency (see, for example,
Ref.\ \cite{krs2}). The most dominant underlying theme of these
models has been the multifractal formalism \cite{Frisch1995}.

There is a similar intermittency in the dissipative scales. This,
too, has been experimentally characterized since Ref.\ \cite{krs1},
and many models have been developed as well (see, again, \cite{krs2}
for a summary). The relation between the two intermittencies has
been the subject of the so-called refined similarity hypothesis
(RSH) put forth in \cite{k62}. This hypothesis links the statistics
of the velocity increments at inertial scales with that of the
velocity gradients at smaller scales where inertial and viscous
ranges match.

In Refs.\ \cite{krs1,Paladin1987}, it was recognized that
dissipation intermittency implies an infinite number of dissipative
scales, $\eta$. Using this insight, Nelkin \cite{Nelkin1990} worked
out the Reynolds number dependence of the moments of velocity
derivatives. Frisch and Vergassola \cite{Frisch1991} denoted the
range of dissipation scales spanning between $\eta_{min}$ and
$\eta_{max}$ as the intermediate dissipation range. Their geometric
picture of the continuum of dissipation scales is that each element
of the range would possess a local H\"older exponent $h$, which
characterizes the spatial roughness of subsets of velocity
increments in the inertial range. Consequently, the minimum and
maximum values of the dissipation scale would be controlled by the
smallest and the largest H\"older exponents:
$\eta_{min}=\eta(h_{min})$ and $\eta_{max}=\eta(h_{max})$. Later,
Chevillard {\it et al.} \cite{Chevillard2005} studied the
intermediate dissipation range within a random cascade model that
takes $\delta_r u$ as a product of a Gaussian random variable and a
positive (scale-dependent) transfer variable. They found that
$\ln(\eta_{max}/\eta_{min})\sim \sqrt{\ln Re}$. The relation of the
intermediate dissipation range to the decay of energy spectra was
discussed recently in the context of well-resolved shell models
\cite{Bowman2006}.

Efforts have also been made to obtain $\zeta_n(n)$ directly from the
Navier-Stokes equations but the problem has remained a great
challenge. We limit ourselves here to citing the work of
\cite{Yakhot2001}---in part because of the connection to the present
work and in part because the author kept his considerations close to
the dynamical equations. The theory has been extended
\cite{yakhot1,yakhot3} to explore the connection between the viscous
and inertial range intermittencies. This extension builds on the
notion that the fluctuating dissipation scale $\eta$ is to be
considered a field that varies in space and time. A relevant feature
of the theory is its prediction for the scaling of velocity
gradients in terms of the exponents $\zeta_n$.

Within this overall framework, the present paper accomplishes the
following goals. First, we perform direct numerical simulations
(DNS) of homogeneous isotropic turbulence with vastly better
spectral resolution than anytime earlier. As was explicitly stated
by Nelkin \cite{Nelkin1990}, such superfine resolutions are required
to compute the derivatives accurately. Second, we then study the
relation between the inertial and dissipative regions within the
framework of existing theories, namely RSH and theory of
\cite{yakhot1,yakhot3}. After describing the theoretical basis
\cite{yakhot1,yakhot3} and the details on the numerical simulations
(Sec.\ 1), we discuss the analyticity and scaling of the velocity
increment moments in Sec.~2 and present our findings on the velocity gradient
statistics in Sec.~3. We also compare in Sec.~3 our results with those of
previous work, study in detail the crossover of the statistics of
the velocity gradients from Gaussian to non-Gaussian regime and
discuss our results in the light of the multifractal formalism.

Perhaps the most surprising result of the present work is that,
while we find no evidence for the inertial range in the DNS data
(even when examined through the extended self-similarity, or ESS),
the measured scaling exponents of velocity gradients agree well with
those deduced from the longitudinal structure functions at {\em
infinitely high Reynolds numbers}. This suggests that the asymptotic
state of turbulence is attained for the velocity gradients at far
lower Reynolds numbers, well short of those required for the
inertial range to appear. We conclude with a summary and an outlook
in Sec.~4.

%----------------------------------------------------------------------------------------------
\begin{table}
\caption{\label{tabone}Parameters of the direct numerical
simulations. Here, $\nu$ is the kinematic viscosity, $\overline{\cal
E}$ is the mean energy dissipation rate,
$R_{\lambda}=\sqrt{5/(3\overline{\cal E}\nu)} u_{rms}^2$ is the
Taylor-microscale Reynolds number. We will use the following
definition for the large scale Reynolds number: $Re =
u_{rms}L_x/\nu$ where the box size $L_x=2\pi$ is taken.
$u_{rms}=(\overline{u_x^2}+\overline{u_y^2}+\overline{u_z^2})^{1/2}$
is used instead of $(\delta_Lu)_{rms}$. The spectral resolution is
indicated by $k_{max}\eta_K$ where $k_{max}=\frac{\sqrt{2}N}{3}$ and
$N$ the number of grid points in each direction of the cube.}
\begin{indented}
\lineup
\item[]\begin{tabular}{@{}*{9}{l}}
\br
Run No.&$N$&$\nu$&$\overline{\cal E}$&$u_{rms}$&$L$&$R_{\lambda}$&$Re$&$k_{max}\eta_K$\cr
\mr
1   & 512  & 1/30  & 0.1  & 0.687 & 1.018 & 10    & 129   & 33.56 \cr
2   & 1024 & 1/75  & 0.1  & 0.808 & 0.920 & 24    & 380   & 33.56 \cr
3   & 1024 & 1/200 & 0.1  & 0.854 & 0.758 & 42    & 1074  & 15.93 \cr
4   & 1024 & 1/400 & 0.1  & 0.892 & 0.694 & 65    & 2243  &  9.6  \cr
5   & 2048 & 1/400 & 0.1  & 0.882 & 0.690 & 64    & 2218  & 19.2  \cr
6   & 2048 & 1/1000 & 0.1 & 0.911 & 0.659 & 107   & 5725  &  9.6  \cr
\br
\end{tabular}
\end{indented}
\end{table}
%-------------------------------------------------------------------------------------------------
\subsection{Numerical Simulations}
The Navier-Stokes equations for an incompressible Newtonian fluid
${\bf u}({\bf x},t)$ are solved in a periodic box of side length
$L_x=2\pi$. The pseudospectral method is applied with a 2/3
de-aliasing for the fast Fourier transforms. Advancement in time is
done by a second order predictor-corrector scheme. The equations are
given by
\begin{equation}
\frac{\partial {\bf u}}{\partial t} + ({\bf u\cdot\nabla}){\bf u} = - {\bf \nabla}p +
\nu \Delta {\bf u}+ {\bf f}\,.
\label{nse}
\end{equation}
The kinematic pressure field is $p({\bf x},t)$ and $\nu$ is the
kinematic viscosity. We consider flows that are sustained by a
volume-forcing ${\bf f}({\bf x},t)$ in a statistically stationary
turbulent state. This driving is implemented in the Fourier space
for the modes with the largest wavenumbers $k_f$ only, i.e.
$k^{-1}_f\approx L_x$. The kinetic energy is injected at a fixed
rate $\epsilon_{in}$ into the flow. The volume forcing is
established by the expression \cite{Eggers1991,Schumacher2004}
\begin{equation}
{\bf f}({\bf k},t)=\epsilon_{in} \frac{{\bf u}({\bf
k},t)}{\sum_{{\bf k}_f\in K} |{\bf u}({\bf k}_f,t)|^2}
\,\delta_{{\bf k},{\bf k}_f}\,, \label{fordef}
\end{equation}
where the wavevector subset $K$ contains ${\bf k}_f=(1,1,2)$ and
(1,2,2) plus all permutations with respect to components and signs.
This energy injection mechanism prescribes the mean energy
dissipation rate; that is, the magnitude of the first moment of the
energy dissipation rate field, $\overline{\cal E}$, is determined by
the injection rate, $\epsilon_{in}$, having no Reynolds number
dependence. This can be seen as follows. Given the periodic boundary
conditions in our system, the turbulent kinetic energy balance,
which results from rewriting (\ref{nse}) in the Fourier space, reads
as:
\begin{equation}
\frac{d E_{kin}}{dt}=-\nu \sum_{\bf k} k^2 |{\bf u}({\bf k},t)|^2 +
\sum_{\bf k} {\bf f}({\bf k},t) {\bf \cdot u}^{\ast}({\bf k},t)\,.
\label{balance}
\end{equation}
The first term on the right hand side of (\ref{balance}) is the
volume average of the energy dissipation field. Additional time
averaging in combination with (\ref{fordef}) results in
\begin{equation}
\nu \sum_{\bf k} k^2 \langle|{\bf u}({\bf k},t)|^2\rangle_t=\overline{\cal E}=\epsilon_{in}=
\sum_{\bf k} \langle{\bf f}({\bf k},t) {\bf \cdot u}^{\ast}({\bf k},t)\rangle_t\,.
\label{epsiloninput}
\end{equation}
The applied driving thus allows a full control of the Kolmogorov
scale $\eta_K=\nu^{3/4}/\overline{\cal E}^{1/4}$ in comparison to
the grid spacing. In contrast to the usually applied stochastic
forcing, the integral length scale $L$, which is defined
\cite{Pope2000} as
\begin{equation}
L=\frac{\pi}{2\,\overline{u_x^2}}\,\int_0^{\infty}\mbox{d}k\,\frac{E(k)}{k}\,,
\end{equation}
decreases with increasing Reynolds number. Since the forcing scale
in the computations is proportional to the box size, we use the box
size $L_x$ as the relevant scale. The use of the integral scale
instead of $L_x$ does not alter the scaling results significantly.
Further details on the simulation parameters can be found in
table~\ref{tabone}.

\begin{figure}
\hspace{4cm}
\includegraphics[scale=0.65]{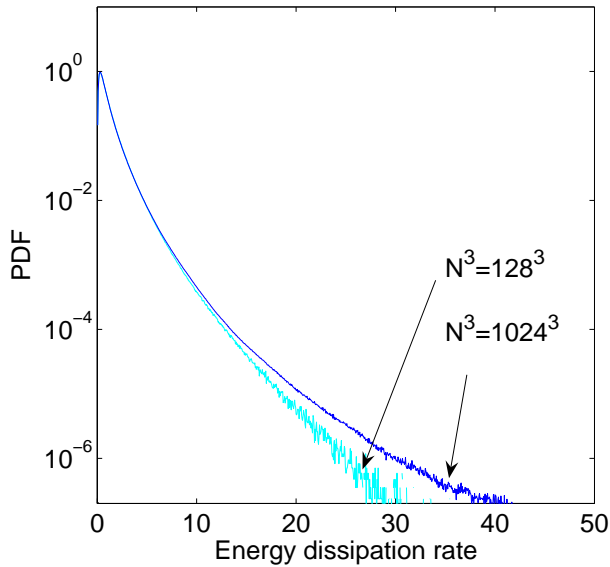}
\caption{Resolution requirements in the numerical simulations. The
probability density function (PDF) of the energy dissipation field
${\cal E}({\bf x},t)=\frac{\nu}{2}(\partial u_i/\partial
x_j+\partial u_j/\partial x_i)^2$ is plotted. The dissipation field
is given in units of the mean energy dissipation rate
$\overline{\cal E}$. The case with $k_{max}\eta_K = 1.2$ (cyan
curve), corresponding roughly to the standard resolution in a box of
size $N=128$, is compared with that of superfine resolution (blue
curve, see also table~\ref{tabone}). While the cores of both PDFs
agree, deviations are manifest in the far tails. Both runs are for
$R_{\lambda}=65$. Approximately $1.7\times 10^8$ data points were
processed for the analysis in the low-resolution run; the
corresponding number for the high-resolution run was about 30 times larger.} 
\label{tails}
\end{figure}

As already pointed out, the adequate resolution of the analytic part
of structure functions turns out to be very demanding. The
resolution in the present simulations exceed those of conventional
simulations by a factor $O(10)$. Consequently, the Reynolds numbers
attained are modest despite the relatively large computational box,
very much in the spirit of \cite{Chen1993}. In order to stress this
point further, we compare the statistics of the energy dissipation
field ${\cal E}({\bf x},t)=\frac{\nu}{2}(\partial u_i/\partial
x_j+\partial u_j/\partial x_i)^2$ for our resolution and the
standard case (see figure~\ref{tails}). At least for the intended
analysis of higher-order gradient statistics the proper resolution
of the far tails turns out to be necessary.
%------------------------------------------------------------------------
%------------------------------------------------------------------------
\begin{figure}
\includegraphics[scale=1.4]{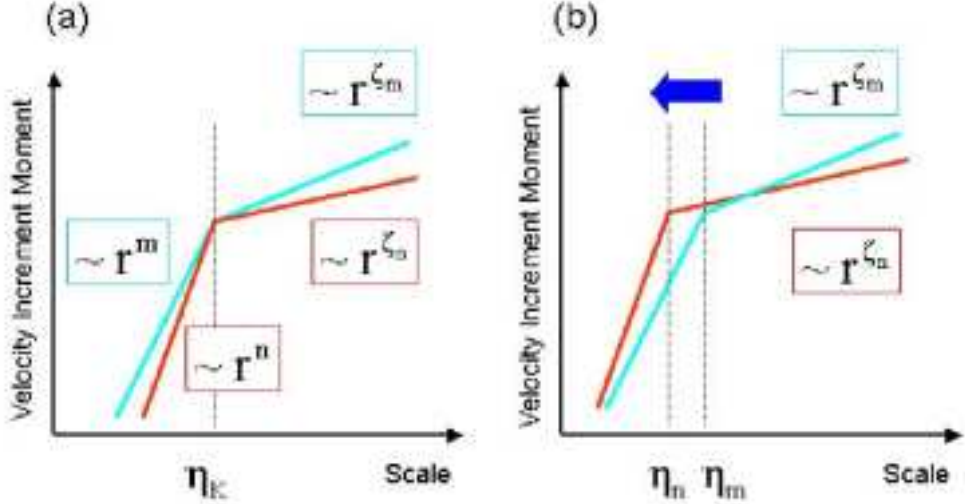}
\caption{The matching of the inertial and dissipative ranges. (a) In
the standard approach, the same Kolomogorov scale matches the
singular ($\sim r^{\zeta_n}$) and analytic parts ($\sim r^n$) for
all orders $n$ of the increment moments. (b) In the theory of
\cite{yakhot1,yakhot3}, for each order of the moment $S_n(r)$, the
singular and analytic parts match at an order-dependent scale
$\eta_n$. Following (\ref{etan}), $\eta_n<\eta_m$ when $n>m$.}
\label{sketch}
\end{figure}
%------------------------------------------------------------------------
\subsection{Theoretical Basis}
The theory \cite{yakhot1,yakhot3} starts with the exact equations
for the $n$-th order longitudinal structure functions $S_{n}(r)$
which can be directly derived from the equations of motion for the
turbulent fluid \cite{Yakhot2001,Hill2001}. For homogeneous,
isotropic and statistically stationary turbulence in three
dimensions these equations take the form
\begin{equation}
\frac{\partial S_{2n}(r)}{\partial r}+\frac{2}{r}S_{2n}(r) =
\frac{2(2n-1)}{r}G_{2n-2,2}(r)+ (2n-1)\overline{\delta_r a(\delta_r
u)^{2n-2}}\,. \label{moments}
\end{equation}
Here, $G_{2n-2,2}$ is the mixed term containing longitudinal
increments of order $2n-2$ and transverse increments of order 2.
Equation (\ref{moments}) is not closed because the last term on the
right hand side is unknown. For small increment scales around the
Kolmogorov length, it follows that the expression for the Lagrangian
acceleration of fluid particles is given by
\begin{equation}
\delta_{\eta} a=\frac{(\delta_{\eta} u)^3}{\nu}\,,
\end{equation}
recalling that the characteristic time is of the order
$\nu/(\delta_{\eta} u)^2$. In this equation, note that $\eta$ is a
field and that increments are therefore taken across variable
distances. The unknown term has the form
\begin{equation}
\overline{\delta_r a(\delta_r u)^{2n-2}} \approx
\frac{1}{\nu}\overline{(\delta_{\eta} u)^3(\delta_r u)^{2n-2}}\,.
\end{equation}
This correlation involves two scales---a locally varying dissipation scale
field $\eta$ in the acceleration increment and $r$ in the velocity
increment moment---and is therefore hard to manipulate. However, in
the limit $r\to\eta$ one can make some progress. In this limit, we
set $\eta=\eta_{2n}$, where $\eta_{2n}$ is the {\em order-dependent}
matching distance between the analytic and singular parts of
$S_{2n}(r)$ (see figure \ref{sketch}). The other three terms of
(\ref{moments}) are found to be of the same magnitude
\cite{Kurien2001,Gotoh2003} and are of the order $S_{2n}/\eta_{2n}$.
One thus obtains, for $r\to\eta$, the result
%----------------------------------------------------------
\begin{equation}
\frac{S_{2n}(\eta_{2n})}{\eta_{2n}}\approx\frac{S_{2n+1}(\eta_{2n})}{\nu}\,.
\label{etan0}
\end{equation}
%----------------------------------------------------------
Now, the velocity increments have the property that, at the large
scale $L$, their distribution is Gaussian. It then follows from (1)
that $S_{2n}(\eta_{2n})=(2n-1)!! \sigma_L^{2n}
(\eta_{2n}/L)^{\zeta_{2n}}$, where
$\sigma_L=\sqrt{\overline{(\delta_L u)^2}}$. \footnote{For example,
$S_4(L)=3!!\sigma^4_L(L/L)^{\zeta_4}=3\sigma^4_L$ and
$S_2(L)=\sigma^2_L$. It follows that, at the appropriate large scale
$L$, the flatness factor $S_4(L)/S_2^2(L)=3$.} Putting $r =
\eta_{2n}$, one obtains from (5)
%----------------------------------------------------------
\begin{equation}
\left(\frac{\eta_{2n}}{L}\right)^{\zeta_{2n}-\zeta_{2n+1}-1}=
\frac{(2n)!!}{(2n-1)!!} \frac{\sigma_L L}{\nu}\approx \frac{\sigma_L L}{\nu}\,.
\label{etan1}
\end{equation}
%----------------------------------------------------------
With the large scale Reynolds number $Re=\sigma_L L/\nu$ one gets
%----------------------------------------------------------
\begin{equation}
\eta_{2n} \approx L Re^{\frac{1}{\zeta_{2n}-\zeta_{2n+1}-1}}\,.
\label{etan}
\end{equation}
%----------------------------------------------------------
For the Kolmogorov scaling, $\zeta_n=\frac{n}{3}$ and (\ref{etan})
yields $\eta_{2n}=L Re^{-3/4}=\eta_K$ for all orders $n$, as
consistency would require.

To make further progress, the functional dependence of $\zeta_{2n}$
has to be given explicitly. The theory of \cite{Yakhot2001} provides
a convenient functional form
%----------------------------------------------------------
\begin{equation}
\zeta_{2n}=\frac{2(1+3\beta)n}{3(1+2\beta  n)},
\label{exponent}
\end{equation}
%----------------------------------------------------------
which, with the free fitting parameter $\beta$ set to 0.05, agrees
with available experimental data in high-Reynolds-number flows (for
order 10-15). We find that this relation agrees, up to order
$10-15$, with available measurements as well as with popular
parametrizations of $\zeta_n$, e.g. with the She-Leveque model and
the $p$-model \cite{She1994,Chen2005}. The scaling behavior of the
spatial derivative in the analytic range of the displacement $r$ can
be calculated, in the limit $r\rightarrow \eta$, as
%----------------------------------------------------------
\begin{equation}
\overline{\left|\frac{\partial u}{\partial x}\right|^{n}}\approx
\overline{\left|\frac{\delta_{\eta} u}{\eta}\right|^{n}}=
\frac{\overline{(\delta_{\eta}u)^{2n}}}{\nu^{n}}\propto
Re^{n}\eta_{2n}^{\zeta_{2n}} = Re^{\rho_{n}},
\end{equation}
%----------------------------------------------------------
where we have used that the ``dynamic" Reynolds number at the local
dissipation scale is unity, i.e.
$Re_{\eta}=\eta\delta_{\eta}u/\nu\approx 1$. The use of (\ref{etan})
yields
%----------------------------------------------------------
\begin{equation}
\rho_{n}=n+\frac{\zeta_{2n}}{\zeta_{2n}-\zeta_{2n+1}-1}\,.
\label{rhon}
\end{equation}
%----------------------------------------------------------
Since $\zeta_3=1$ \cite{k41b}, relation (\ref{rhon}) gives
$\zeta_2=(2-2\rho_{1})/(2-\rho_{1})$. For the Kolmogorov value of
$\zeta_{2}=2/3$, we obtain $\rho_{1}=1/2$. The anomaly that may exist
in the first-order exponent $\rho_{1}$ for velocity gradients is
related to the second-order inertial exponent $\zeta_2$.
For moments of the dissipation rate, one can write
%--------------------------------------------------------
\begin{equation}
\overline{{\cal E}^{n}} \equiv
\nu^{n}\,\overline{\left(\frac{\partial u_{i}}{\partial
x_{j}}\right)^{2n}}\approx
\nu^{n}\,\overline{\left(\frac{(\delta_{\eta}
u)^{2}}{\nu}\right)^{2n}}=
\frac{\overline{(\delta_{\eta}u)^{4n}}}{\nu^{n}}\propto Re^{d_{n}},
\end{equation}
%--------------------------------------------------------
where
%--------------------------------------------------------
\begin{equation}
d_{n}=n+\frac{\zeta_{4n}}{\zeta_{4n}-\zeta_{4n+1}-1}.
\label{dn}
\end{equation}
%--------------------------------------------------------
Thus, the second-order quantities for the dissipation rate are
expressed in terms of the eighth-order quantities involving velocity
increments. In general, to accurately evaluate $\overline{{\cal
E}^{n}}$, one has to resolve the analytic range within which
$S_{4n}(r)\propto r^{4n}$. This difficulty for large $n$ is one of
the main considerations of the theory.

Finally, we note that
\begin{equation}
d_1=0\
\end{equation}
because $\overline{\cal E}=\epsilon_{in}$ holds in the DNS (see
equation~(\ref{epsiloninput})). From relation (\ref{dn}) one immediately
has
\begin{equation}
\zeta_5=2\zeta_4-1\,.
\end{equation}

Measurements are, in fact, in good conformity with this equation.

\subsection{Velocity Gradients from Refined Similarity Hypothesis}
The refined similarity hypothesis (RSH) \cite{k62} imposes a
different constraint between moment orders of the energy dissipation
and structure functions. This results in a different Reynolds number
dependence compared to (\ref{dn}). When taking ${\cal
E}\sim(\delta_{\eta_K}u)^3/\eta_K$ the relation
\begin{equation}
\overline{{\cal E}^n}\sim \frac{S_{3n}(\eta_K)}{\eta_K^n}
\sim \eta_K^{\zeta_{3n}-n}\,,
\end{equation}
follows. With $\eta_K=L Re^{-3/4}$ we get
\begin{equation}
\overline{{\cal E}^n}\sim Re^{\frac{3(n-\zeta_{3n})}{4}}\,.
\label{rsh}
\end{equation}
The comparison with the present data will be made later in this text
(see table~\ref{tabtwo}). Here, we briefly mention that the
intermittency exponent $\mu$ in the scaling $\overline{{\cal E}({\bf
x}+{\bf r}){\cal E}({\bf x})}\sim r^{-\mu}$ is 0.25 from RSH while
the application of the theory of Refs.\ \cite{yakhot1, yakhot3}
gives $\mu\approx 0.2$. Both are within the accepted range of $0.25
\pm 0.05.$

\subsection{Velocity Gradients from the Multifractal Formalism}
Nelkin \cite{Nelkin1990} predicted a Reynolds number dependence for
derivative moments based on the multifractal formalism (MF). The
derivation of his expressions relied on the steepest descent
approximation of resulting integrals and input from measurements
\cite{krs1}. His result (which is also outlined in detail in
\cite{Frisch1995}, pp.\ 157-158) is that
\begin{equation}
\rho_n=p(n)-n\,,
\label{rhonmf}
\end{equation}
in terms of the notation used above. Here $p(n)$ is a unique
solution that follows from the intersection of the concave curve
$\zeta_p$ with the straight line $2n-p$ for a given $n$. With
(\ref{exponent}) the intersection ($p>0$) obeys the relation
\begin{equation}
p(n)\approx-\frac{4+3\beta(1-2n)}{6\beta}+\frac{1}{6\beta}\sqrt{72\beta
n+(3\beta(2n-1)-4)^2}\,. \label{nullstelle}
\end{equation}
The resulting values of $\rho_n$ are also listed in
table~\ref{tabtwo}.

\section{Moments of Velocity Increments}
\subsection{Rescaling of Higher Order Moments and Test of Analyticity}
%------------------------------------------------------------------------
\begin{figure}
\hspace{1cm}
\includegraphics[scale=0.75]{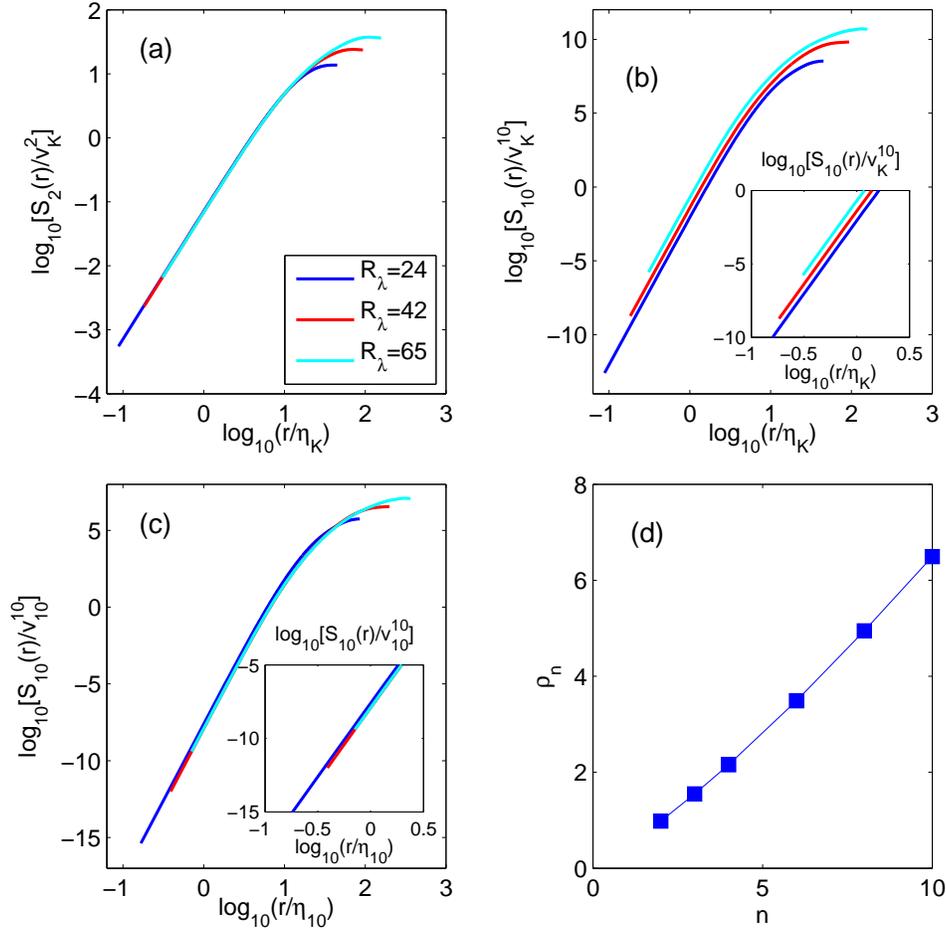}
\caption{Longitudinal structure functions of the turbulent velocity
field. (a) Second-order longitudinal structure function
$S_{2}(r)/v_{K}^{2}$ over $r/\eta_K$ for the three different Runs
indicated in the legend. (b) Tenth-order longitudinal structure
functions $S_{10}(r)/v_{K}^{10}$ over $r/\eta_K$ for the same data.
The inset shows that the curves do not collapse well. Here,
$v_{K}=(\nu \overline{\cal E})^{1/4}$ is the Kolmogorov velocity.
(c) The tenth-order structure functions for $R_{\lambda}=24,42,65$
collapse when $r$ is rescaled by the dissipation scale $\eta_{10}$
defined by (\ref{etan}) and the amplitudes by the velocity scale
$v_{10}^{10}$ (see equation (\ref{vetan})). The inset shows the same
level of expansion as in (b). (d) The figure shows the exponent
$\rho_n$ as a function of order $n$ given by
equations~(\ref{exponent}) and (\ref{rhon}).} \label{fig1}
\end{figure}
%------------------------------------------------------------------------
\begin{figure}
\centerline{\includegraphics[angle=0,scale=0.6,draft=false]{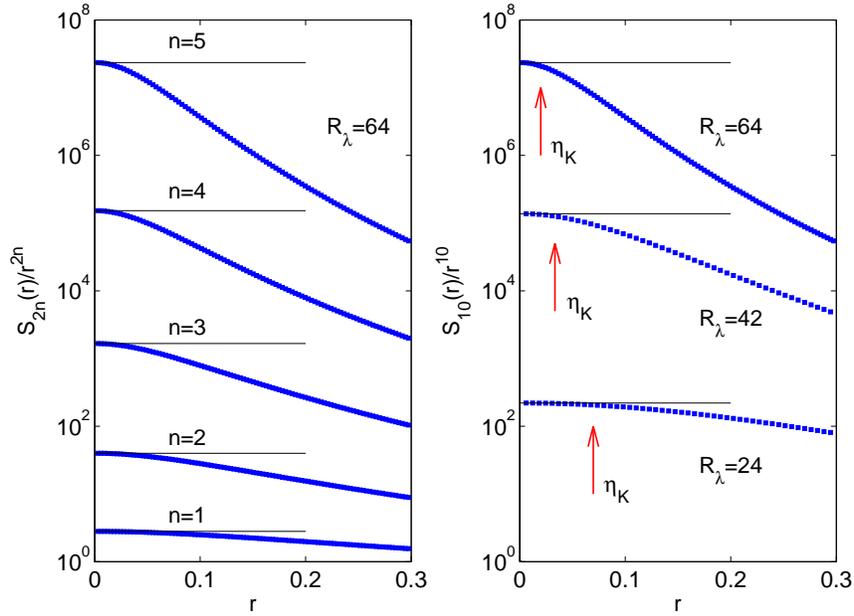}}
\caption{Test of analyticity for the longitudinal structure
functions $S_{2n}(r)$ by compensated plots. Left: Orders
$2n=2,4,6,8,10$ are shown for Run 5. Horizontal lines indicate the
exact analytical form corresponding to $S_{2n}(r)\sim r^{2n}$.
Right: $S_{10}(r)$ is shown for Runs 2, 3 and 5. The red vertical
arrows indicate the corresponding Kolmogorov scales $\eta_K$.}
\label{fig2}
\end{figure}
%-------------------------------------------------------------------------------
Figures~\ref{fig1}a and \ref{fig1}b show the structure functions
$S_{2n}(r)/v_{K}^{2n}$ plotted against $r/\eta_K$ for three Reynolds
numbers for order 2 $(n=1)$ and 10 $(n=5)$. Here, $v_K$ is the
Kolmogorov velocity corresponding to the definition $\eta_K v_K/\nu
= 1$. Two features of the graphs are worth noting. First, no
inertial range can be seen at these low Reynolds numbers. Second,
all data possess the analytic parts thus confirming that the
resolution used is adequate. It is important to recognize in
figure~\ref{fig2} that with increase of both the Reynolds number and
moment order the width of the analytic range decreases. To make this
point more explicit, we plot in figure~\ref{fig2} the compensated
structure functions $S_{2n}/r^{2n}$ ($n=1-5$) for a fixed Reynolds
number $R_{\lambda}=64$ (left) and the normalized moment
$S_{10}/r^{10}$ for different Reynolds numbers (right). In the
analytic range we expect
\begin{equation}
\frac{S_{2n}(r)}{r^{2n}}\rightarrow\overline{\left(\frac{\partial
u}{\partial x}\right)^{2n}}=const\,,
\end{equation}
i.e., independent of the increment distance $r$. For
$R_{\lambda}=64$, this range is well-defined for the moments $n\leq
3$ and just survives for $n=5$. This means that, even for the
present super-resolution, the representation of moments of velocity
derivatives in terms of the low-order finite differences may be
problematic for moment orders higher than 5 (although we do present
the data for $n$ up to 8).

In the analytic range, all curves can be expected to collapse when
normalized by the appropriate length and velocity scales. The
traditional scales are the Kolmogorov length and velocity scales
$\eta_K$ and $v_{K}$, respectively. This scaling works well for 
low order moments (say 2), as seen in figure~\ref{fig1}a. The same
normalization is not adequate for high orders such as 10, as can be
seen in figure~\ref{fig1}b. However, all curves do collapse when the
length scale $\eta_{2n}$ (see equation (\ref{etan})) and the
corresponding velocity scale
\begin{equation}
v_{2n}=\frac{\nu}{\eta_{2n}}
\label{vetan}
\end{equation}
are used instead of $\eta_{K}$ and $v_K$ (see
figure~\ref{fig1}c).\footnote{We note here parenthetically that
$S_{10}$ for the lowest Reynolds number ($R_\lambda = 10$) does not
collapse on the common curve. For this case, however, the velocity
gradient statistics are Gaussian, in contrast to the other three
cases. This issue will be discussed in greater detail in Sec.\ 3.2.}

How can we understand this collapse? Introducing the ``dynamic"
Reynolds number $Re_{r}=r\delta_{r}u /\nu$, and recalling that the
only relevant parameter in the inertial range---including the
interval just above the dissipation range ($r \gtrsim \eta$)---is
the energy flux, we can conclude that the dynamics of fluctuations
at the scales $r\approx \eta$ are independent of the width of the
inertial range $(L,\eta)$. While one cannot, in principle, rule out
the existence of other fluxes corresponding to some yet unknown
integrals of motion related to structure functions of higher order
than 2 \cite{fs}, this possibility will not influence the essence of
the argument. If so, for $r\lesssim \eta$, the properly normalized
moment of a given order $n$ must be independent of $Re$.

\subsection{Test of Extended Self-Similarity}
%------------------------------------------------------------------------
\begin{figure}
\hspace{2cm}
\includegraphics[scale=0.75]{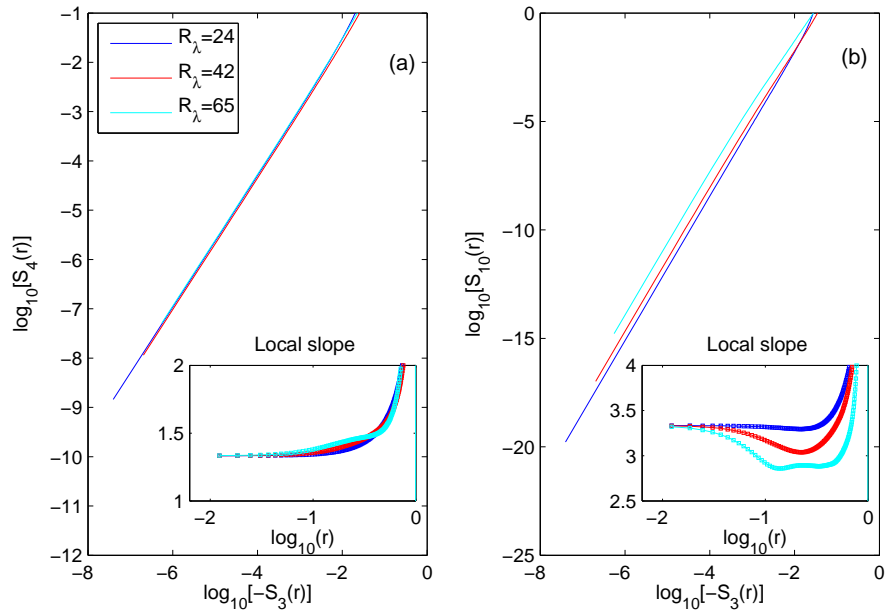}
\caption{Extended self-similarity analysis of longitudinal structure
functions. (a) Fourth-order structure function $S_4(r)$ versus
$-S_3(r)$. Structure functions are defined by
equation~(\ref{structuref}), with $-S_3(r) = (4/5)\overline{\cal E} r$. (b) Tenth-order
structure function $S_{10}(r)$ versus $-S_3(r)$. The insets in both
figures show the corresponding local slope $\chi_n(r)$ as given by
relation (\ref{localslope}). Data of Runs 2,3 and 5 are indicated by
different colours, as noted in the legend.} \label{esscheck}
\end{figure}
%------------------------------------------------------------------------
Since no inertial-range scaling can be observed for such low
Reynolds numbers in the standard double logarithmic plot, we tested
the method of extended self-similarity (ESS) by Benzi {\it et al.}
\cite{Benzi1993}. As is well known, the main point of ESS is that
even though there is no discernible scaling in the standard plot
when the Reynolds numbers are low, one can detect a sizeable range
of scaling when arbitrary moment orders are compared with $S_3$:
%----------------------------------------------------------------------
\begin{equation}
S_n(r)\sim (-S_3(r))^{\zeta_n}\,.
\end{equation}
%----------------------------------------------------------------------
The sensitive and scale-dependent measure for possible anomalous
scaling is then the local slope
\begin{equation}
\chi_n(r)=\frac{\mbox{d}\ln(S_{n}(r))}{\mbox{d}\ln(-S_3(r))}\,.
\label{localslope}
\end{equation}
The results for our data are shown in figure~\ref{esscheck} for the
fourth and tenth order. In both figures we cannot detect anomalous
scaling in the ESS framework, which would result in local slopes of
$\chi_4(r)\approx 1.28$ and $\chi_{10}(r)\approx 2.59$
\cite{She1994,Yakhot2001}. We conclude that only a further increase
of the Reynolds number will shift the local slope toward the
asymptotic values. Thus, even the ``backdoor" of ESS is not opened
for our small Reynolds numbers. It confirms our statement made
before that the velocity field statistics does not have any
asymptotic scaling in the inertial range. The inset reveals an
interesting feature. It can be observed that a faster relaxation
towards anomalous scaling occurs with increasing order. At this
point, we can only speculate about the reason for this feature. On
one hand, the differences between the viscous scaling,
$\chi_n(r)=n/3$, and the inertial scaling, $\chi(r)=\zeta_n$, become
larger with increasing order and thus better visible in the local
slope. On the other hand, an order-dependence of the dissipation
scale, $\eta_n$, might cause a slight increase of the inertial range
and the crossover between inertial and viscous ranges, respectively.

\section{Velocity Gradient Statistics}
\subsection{Results and Comparison with Refined Similarity Hypothesis}
If indeed the properties of fluctuations from the interval
$r\lesssim \eta$ depend only upon the local magnitude of the energy
flux ${\overline {\cal E}}$ and {\it not} upon the width of the
inertial range $(L,\eta)$, one can hope to obtain the asymptotic
values of exponents $\rho_{n}(Re)\rightarrow \rho_{n}(\infty)$ in
reasonably low-Reynolds-number flows, provided that a small (even
very small) constant-flux range exists for scales $r>\eta$.

\begin{figure}
\centerline{\includegraphics[angle=0,scale=0.22,draft=false]{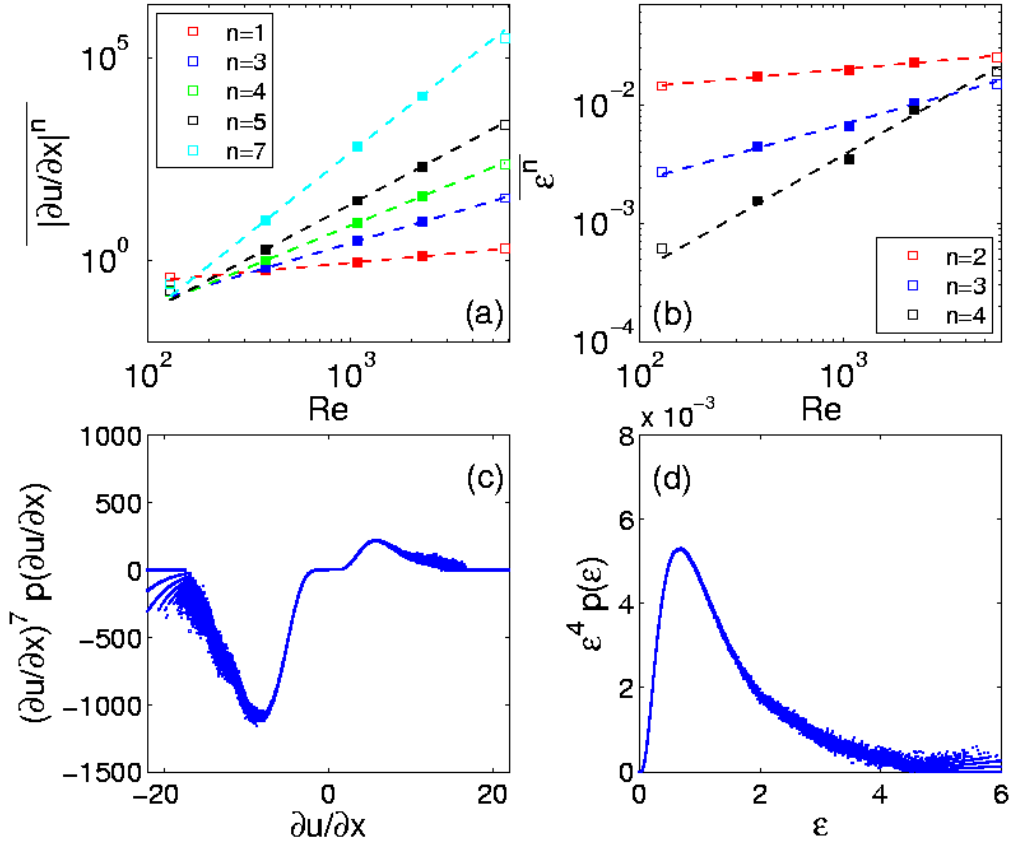}}
\caption{The Reynolds number dependence of the moments of velocity
derivative and energy dissipation. (a) Moments of the absolute value
of the longitudinal velocity derivative $\partial u/\partial x$ for
orders 1, 3, 4, 5 and 7 as functions of the box Reynolds number
$Re$. Only the filled data points were included in the least square
fit, but Run 6 demonstrates that the scaling continues for higher
Reynolds number. The case with the lowest Reynolds number has
Gaussian statistics (as will be described below) and is hence not
turbulent in the traditional sense. (b) Moments of order 2, 3 and 4
of the energy dissipation field ${\cal E}$. Again, only the filled
data points were used for the fit to evaluate the exponent. (c)
Statistical convergence test for the seventh order longitudinal
velocity derivative moment at the highest Reynolds number considered
here, $R_{\lambda}=65$ (Run 4). The data set contained 15 samples of
the turbulent field which results in $1.6\times 10^{10}$ data
points. (d) Statistical convergence test of the fourth order moment
of the energy dissipation field. Again, data from Run 4 are used.}
\label{fig3}
\end{figure}

In figure~\ref{fig3}, we plot $\overline{|\partial u/\partial x|}$ and
$\overline{{\cal E}^n}$ as functions of $Re$. Statistical
convergence of both gradient quantities is satisfactory for gradient
moments of at least up to the $7$-th order (see figure~\ref{fig3}c)
and dissipation moments of at least up to the 4-th order (see
figure~\ref{fig3}d). As can be seen from table~\ref{tabtwo}, the data agree well
with theoretical predictions \cite{yakhot1,yakhot3}. For example, we
get $\rho_1=0.455$ compared to the theoretical value of 0.465. The
second order exponent $\zeta_2$ is given by
\begin{equation}
\zeta_2=\frac{2-2\rho_1}{2-\rho_1}=0.706\,,
\end{equation}
which is very close to the experimental value of $\zeta_2=0.71$
\cite{Benzi1993,Chen2005}. We stress that this result was obtained
in flows with $24\leq R_{\lambda}\leq 65$, none of which has any
inertial range. The moments of the dissipation rate also agree with
theoretical predictions, as shown in table~\ref{tabtwo}.
%-------------------------------------------------------------------------------
%-----------------------------------------------------------------------

In the theory outlined earlier \cite{yakhot1}, the second, third and
fourth moments of ${\cal E}$ are related to the structure functions
$S_{8}(\eta_{8})$, $S_{12}(\eta_{12})$ and $S_{16}(\eta_{16})$,
respectively. These structure functions probe very intense,
low-probability velocity fluctuations and, as a result, huge data
sets are needed to accurately evaluate their characteristics. We
have seen that dissipation moments of order 5 are barely resolved in the present
simulations, while the data for the sixth-order moment
$\overline{{\cal E}^{6}}$, corresponding to $S_{24}$ have not
converged well. In addition, we stress that the statistical
convergence is not sufficient for the accurate determination of
structure functions: the simulations must resolve accurately at
least a fraction of the analytic range $r < \eta_{4n}$.

Table 2 lists comparisons between theoretical considerations and the
DNS data. In case of the dissipation field, we inserted
(\ref{exponent}) into (\ref{rsh}); for the velocity derivative
scaling exponents from multifractal formalism, we used the relation
(\ref{nullstelle}) which was inserted into (\ref{rhonmf}). For the
gradient exponents, the DNS data are somewhat smaller than both the
theory \cite{yakhot1,yakhot3} and the multifractal theory. In the
case of the dissipation exponents, the DNS results are closer to the
theory of Refs.\ \cite{yakhot1,yakhot3} and somewhat larger than
those of RSH. Note that the results of both theories depend on the
inertial scaling exponents obtained from measurements (or at least
expressions tuned to agree with measurements). Considering that
there are issues of resolution in measurement (although not in the
same sense as in simulations), these departures may suggest that
inertial exponents may need slight revision. Of course, there might
be other reasons for these differences.

%-------------------------------------------------------------------------------
\begin{table}
\caption{\label{tabtwo}Comparison of scaling exponents for different
velocity gradient moments. $\rho_n$ for $n=1,3,5,7$ (see
(\ref{rhon})) and $d_n$ (see (\ref{dn})) for $n=1,2,3,4$ are listed.
Results from the present DNS are compared with those from the theory
\cite{yakhot1,yakhot3} after inserting (\ref{exponent}) into
(\ref{rhon}) and (\ref{dn}). Comparisons with the refined similarity
hypothesis (RSH) and the multifractal formalism (MF) are also
provided. The error bars for orders 3 and 4 of the DNS data have
been determined from lower and upper envelopes to the tails of
${\cal E}^n p({\cal E})$, $p({\cal E})$ being the probability
density function of the energy dissipation field. The same holds for
orders 5 and 7 in case of $p(|\partial u/\partial x|)$. The range of
Reynolds numbers for all fits was 380 to 2243.}
\begin{indented}
\lineup
\item[]\begin{tabular}{@{}*{5}{l}}
\br
& Theory \cite{yakhot1}& DNS & RSH & MF\cr
\mr
$\rho_1$ & 0.465 & 0.455  & -- & 0.474\cr
$\rho_3$ & 1.548 & 1.478  & -- & 1.573\cr
$\rho_4$ & 2.157 & 2.051  & -- & 2.188\cr
$\rho_5$ & 2.806 & 2.664 $\pm$ 0.137 & -- & 2.841\cr
$\rho_7$ & 4.203 & 3.992 $\pm$ 0.653 & -- & 4.241\cr\mr
$d_1$    & 0.000 & 0.000  & 0.000 & --\cr
$d_2$    & 0.157 & 0.152  & 0.173 & --\cr
$d_3$    & 0.489 & 0.476 $\pm$ 0.009 & 0.465 & --\cr
$d_4$    & 0.944 & 0.978 $\pm$ 0.034 & 0.844 & --\cr
\br
\end{tabular}
\end{indented}
\end{table}

\subsection{Transition from Gaussian to non-Gaussian Velocity Gradient Statistics}
%------------------------------------------------------------------------
\begin{figure}
\includegraphics[scale=1.2]{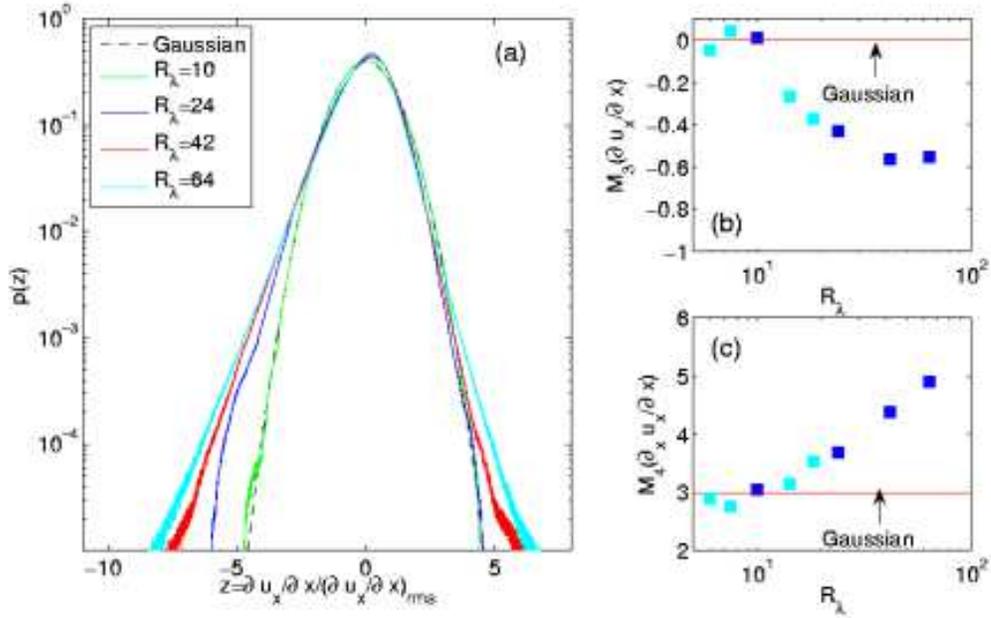}
\caption{Statistics of the velocity gradient $\partial u/\partial
x$. (a) Plots of the probability density functions (PDF) for Runs 1
to 4 as indicated in the legend. The gradient is normalized by its
root-mean-square value. For comparison the Gaussian distribution is
added to the figure. (b) Corresponding skewness of the PDFs (blue
symbols). (c) Corresponding flatness of the PDFs (blue symbols). In
order to highlight the transition, further data points have been
added to the data of table~\ref{tabone} in figures (b) and (c).
These additional numerical simulations have been conducted at a
spectral resolution of $N=256$ for different kinematic viscosities.
Forcing scheme and energy injection rate are the same as before. The
additional data values are plotted as cyan filled squares.}
\label{crossover}
\end{figure}
%------------------------------------------------------------------------
\begin{figure}
\hspace{2cm}
\includegraphics[scale=1.2]{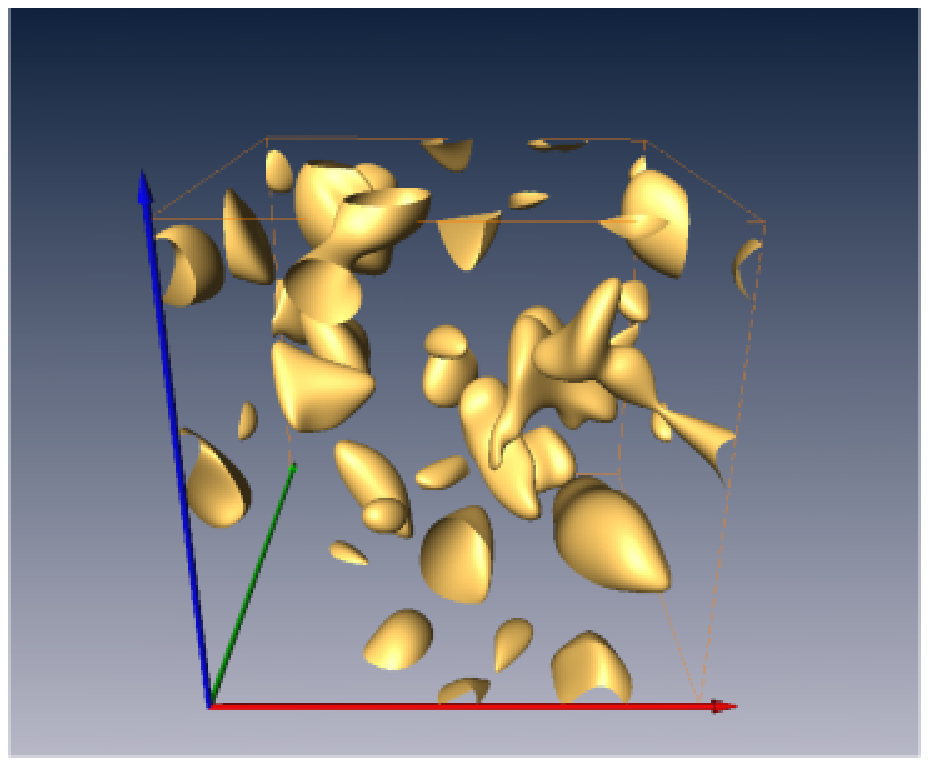}

\hspace{2cm}
\includegraphics[scale=1.2]{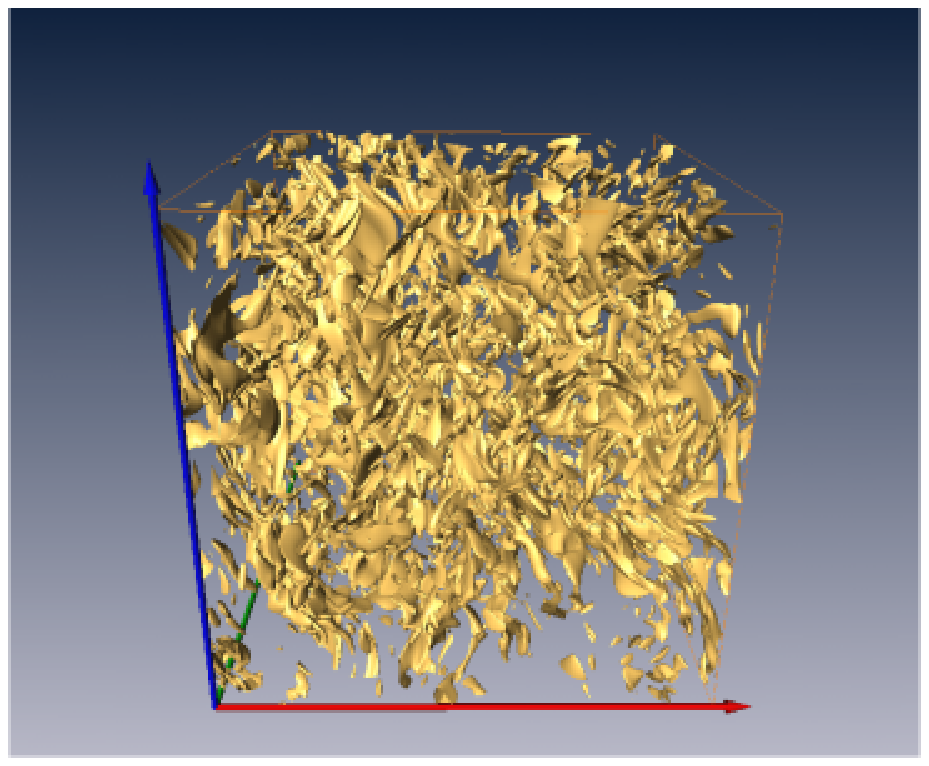}
\caption{Morphological manifestation of the crossover from Gaussian
to non-Gaussian velocity gradient statistics. Isolevel plots of
snapshots of $|\partial u/\partial x|$ for Run 1 at $R_{\lambda}=10$
(top) and for Run 4 at $R_{\lambda}=65$ (bottom) are shown. Both
level sets were taken at the corresponding values of $2\times
(\partial u/\partial x)_{rms}$. We observe a significant increase of
the spatial intermittency for the higher Reynolds number data set. }
\label{isolevel}
\end{figure}
%------------------------------------------------------------------------
As mentioned earlier, while computing scaling exponents, Run 1 was
excluded from the least-square fits. The reason is the qualitatively
different nature of the velocity gradient statistics at this lowest
Reynolds number ($R_{\lambda} = 10$). This will be discussed now.
Figure \ref{crossover} illustrates the crossover from Gaussian
statistics at very low Reynolds numbers to an increasingly
non-Gaussian behaviour for moderate Reynolds numbers. In order to
highlight this transition, we generated additional DNS data at
intermediate Reynolds numbers. The Gaussian values for the third and
fourth order normalized derivative moment are indicated by the red
solid lines in panels (b) and (c) of figure 7. The derivative
moments are defined as
\begin{equation}
M_n(\partial u/\partial x)=\frac{\overline{(\partial u/\partial
x)^n}}{\overline{(\partial u/\partial x)^2}^{\;n/2}}\,.
\end{equation}

For the lowest Reynolds numbers we detected a regime which is a
complex time-dependent flow rather than a turbulent one. In this
regime, the flow can be basically described by the small number of
driven modes that form a low-dimensional nonlinear dynamical system.
All other degrees of freedom are strongly damped and slaved to the
driven modes. The sign of the derivative skewness $M_3(\partial
u/\partial x)$ there became sensitively dependent on the particular
modes that were driven. Since the non-Gaussian behavior is related
to the acceleration-velocity term in equation (7), the experience
suggests that it is difficult to pin down the behavior of this term
near zero. The magnitudes of $M_3$ and $M_4$ varied, respectively,
around zero and three (even for very long-time runs). We verified
this by choosing different wave vectors for driving the flow while
leaving all other simulation parameters the same, including the
energy injection rate $\epsilon_{in}$ (see
equation~(\ref{epsiloninput})). At larger Reynolds numbers,
$R_{\lambda}>10$ or $15$, the derivative moments became insensitive
to the particular driving. Therefore, our studies suggest that the
transition to non-Gaussian statistics is smooth with respect to the
Reynolds number and that intermittency continuously grows with the
growing number of excited modes.

Figure \ref{isolevel} illustrates the morphological changes of the
spatial distribution of the velocity gradient that are connected
with the change of the statistical properties. The increasing
intermittency of the velocity gradients is accompanied by an
increasing fragmentation of the isolevel sets.

\section{Summary and Discussion}
Turbulence in a three-dimensional periodic box, generated by the
Navier-Stokes equations driven by a large-scale forcing, was
investigated in the Taylor microscale Reynolds numbers range $10\leq
R_{\lambda}\leq 107$. The simulations were made with superfine
resolution in order to resolve the analytic part of structure
functions at least up to order 16. No inertial ranges characterized
by the velocity structure functions $S_{n}(r)\propto r^{\zeta_n}$
were detected---not even with the method of extended
self-similarity. In the range $24\leq R_{\lambda}\leq 107$, strong
intermittency of the spatial derivatives was detected and their
moments were accurately described by the scaling relations
$\overline{|\partial u/\partial x|^{n}}\propto Re^{\rho_{n}}$ and
$\overline{{\cal E}^{n}}\propto Re^{d_{n}}$, respectively. The
exponents $\rho_n$ and $d_{n}$ were found to be in essential
agreement with the theoretical work for high Reynolds number
\cite{yakhot1,yakhot3}. Based on the well-resolved numerical
results, we are thus able to relate the dissipation-range exponents
of the velocity gradients to the inertial range scaling exponents of
the velocity field for very high Reynolds numbers. For instance, the
DNS result $\rho_{1}\approx 0.455$ gives $\zeta_{2}=0.706$
corresponding to the asymptotic high-Reynolds number energy spectrum
$E(k)\propto k^{-1.706}$.

The competing theoretical predictions are from RSH (for dissipation)
and the multifractal theory (for the gradients). They, too, agree
with the DNS results, although somewhat less successfully overall.
It appears that the theory of Refs.\ \cite{yakhot1,yakhot3} seems to
have an edge.

This last conclusion would have been more conclusive if we had been
able to obtain reliable dissipation range statistics accurately for
moments of order 5 or 6. Despite the huge data sets of $\approx
10^{10}$ points, we were unable to obtain reliable data for
dissipation field moments with $n > 4$, certainly for $n$ no larger
than 5. The theory suggests the reason for this problem: moments of
the dissipation rate $\overline{{\cal E}^{n}}$ are related to
high-order structure functions $S_{4n}(\eta_{4n})$ probing very low
probability fluctuations evaluated on the corresponding dissipation
scales.

We also conducted simulations at various Reynolds numbers to study
the smooth transition of the velocity gradient statistics from the
Gaussian behaviour to the non-Gaussian behaviour.

Our work suggests that the scaling exponents of the moments of
velocity derivatives observed in the relatively low-Reynolds-number
turbulent flow, lacking even traces of the inertial range, can be
expressed in terms of the inertial-range exponents corresponding to
the {\it asymptotic} case ($Re\rightarrow\infty$). We stress that
the scaling exponents $\rho_{n}$ and $d_{n}$ of the moments of
velocity derivatives and dissipation rate reach asymptotic values
that are independent of large-scale Reynolds number, even at low
values of the Reynolds numbers. Thus, the dynamics of velocity
fluctuations at the scales $r\approx \eta$ are asymptotic even in
relatively low-Reynolds-number flows. This could mean that the
magnitudes of inertial-range exponents ($Re\rightarrow\infty)$ are
prescribed by the matching conditions on the ultra-violet cut-offs
formed in the low-Reynolds-number regimes. The relation of the
observed behavior to fluctuations of the dissipation scale will be
discussed elsewhere.

\ack We thank for supercomputing ressources on the IBM Power 4 cluster
JUMP of the John von Neumann Institute for Computing (NIC) in
J\"ulich (Germany). JS wishes to thank Marc-Andr\'{e} Hermanns (NIC)
for assistance with the parallel performance analysis on the JUMP
system and Michael Rambadt (NIC) for his help with the UNICORE
platform. The supercomputer time on up to 512 CPUs was provided by a
grant of the Deep Computing Initiative of the Distributed European
Infrastructure for Supercomputer Applications consortium (DEISA). We
acknowledge helpful comments and suggestions by R. Benzi, L.
Biferale, L. Chevillard, B. Eckhardt, U. Frisch, T. Gotoh, F. de Lillo and C.
Meneveau.

\end{document}